# Complex Fault Geometry of the 2020 M<sub>WW</sub>6.5 Monte Cristo Range, Nevada Earthquake Sequence

Christine J. Ruhl, Emily A. Morton, Jayne M. Bormann, Rachel Hatch-Ibarra, Gene Ichinose, and Kenneth D. Smith

*Christine J. Ruhl, Corresponding Author*

*Department of Geosciences*

*The University of Tulsa*

*800 South Tucker Drive*

*Tulsa, Oklahoma 74104-9700 USA*

cruhl@utulsa.edu

**Declaration of Competing Interests**

The authors acknowledge there are no conflicts of interest recorded.

**ABSTRACT**

On 15 May 2020 an *M*<sub>WW</sub> 6.5 earthquake occurred beneath the Monte Cristo Range in the Mina Deflection region of western Nevada. Rapid deployment of eight temporary seismic stations enables detailed analysis of its productive and slowly decaying aftershock sequence (p=0.8) which included ~18,000 autodetected events in 3.5 months. Double-difference, waveform-based relative relocation of 16,714 earthquakes reveals a complex network of faults, many of which cross the inferred 35-km long east-northeast-striking, left-lateral mainshock rupture. Seismicity aligns with left-lateral, right-lateral, and normal mechanism moment tensors of 128 of the largest earthquakes. The mainshock occurred near the middle of the aftershock zone at the intersection of two distinct zones of seismicity. In the western section, numerous subparallel, shallow, north-northeast-striking faults form a broad flower-structure-like fault mesh that coalesces at depth into a near-vertical, left-lateral fault. We infer the near-vertical fault to be a region of significant slip in the mainshock and an eastward extension of the left-lateral Candelaria fault. Near the mainshock hypocenter, seismicity occurs on a northeast-striking, west-dipping structure which extends north from the Eastern Columbus Salt Marsh normal fault. Together, these two intersecting structures bound the Columbus Salt Marsh tectonic basin. East of this intersection and the mainshock hypocenter, seismicity occurs in a narrow, near-vertical, east-northeast-striking fault zone through to its eastern terminus. At the eastern end, the aftershock zone broadens and extends northwest towards the southern extension of the northwest-striking, right-lateral Petrified Springs fault system. The eastern section hosts significantly fewer aftershocks than the western section, but has more moment release. We infer that shallow aftershocks throughout the system highlight fault-fracture meshes that connect mapped fault systems at depth. Comparing earthquake data to surface ruptures and a simple geodetic fault model sheds light on the complexity of this recent M6.5 Walker Lane earthquake.

**INTRODUCTION**

The 15 May 2020 $M_{WW}$ 6.5 Monte Cristo Range, Nevada earthquake (11:03:27 UTC) ruptured a complex series of faults in a 35-km-long east-northeast trending zone in the Mina Deflection region of the Central Walker Lane (Figure 1). The earthquake occurred in an uninhabited region of Nevada, but it produced strong to very strong shaking in the nearby communities of Tonopah and Hawthorne, NV, located 55 km and 75 km from the epicenter, respectively. Weak shaking was reported throughout Nevada, and more than 22,000 people submitted "Did You Feel It?" community internet intensity felt reports from regions extending between Salt Lake City, UT (590 km), the San Francisco Bay area (395 km), and San Diego, CA (605 km) (USGS, see Data and Resources). Damage reports are limited to several zones of cracking across US Highway 95 near the epicenter and minor non-structural damage such as broken windows, cracked plaster, and fallen objects in Tonopah, NV. No injuries were reported (Gunay *et al.*, 2020). Within two days of the mainshock, the Nevada Seismological Laboratory (NSL) deployed eight temporary seismic stations which recorded the prolific aftershock sequence (Bormann et al., 2021; Figure 2).

The Monte Cristo Range earthquake is the largest to occur in Nevada since the December 1954 Fairview Peak M 7.1 and Dixie Valley M 6.9 earthquakes and the largest earthquake in the Walker Lane since the July 2019 $M_w$ 6.4 and $M_w$ 7.1 Ridgecrest earthquakes. The Walker Lane (Stewart, 1988) is a 100-to-300-km-wide, northwest-trending zone of discontinuous active faults that straddle the California-Nevada border between the northwest translating Sierra Nevada microplate and the westward extending Basin and Range tectonic province (Figure 1). Faults in the Walker Lane collectively accommodate up to 20% of the 50 mm/yr of relative right-lateral motion between the Pacific and North American plates (e.g., Thatcher *et al.*, 1999; Dixon *et al.*, 2000; Bennett *et al.*, 2003). In the Central Walker Lane, this northwest-directed right-lateral strain is accommodated through a combination of three structural mechanisms: (1) left-lateral slip and clockwise block rotations between east-west-striking faults; (2) right-lateral slip and block translation on northwest-striking faults; and (3) extension on north-south-striking normal faults (see e.g., Wesnousky, 2005).

The structural complexity of the Walker Lane exists on many spatial scales - including the scale of the Monte Cristo Range earthquake. The NSL catalog location for the $M_{WW}$ 6.5 mainshock epicenter lies on the western flank of the Monte Cristo Range, near the intersection between the east-northeast-striking, left-lateral Candeleria fault (Wesnousky, 2005) and the north-striking, westward-dipping Eastern Columbus Salt Marsh normal fault (USGS, 2020). The east-west aftershock zone suggests that the mainshock ruptured an eastward extension of the mapped trace of the Candelaria fault zone. NSL catalog locations of the aftershock sequence span the transition between the east-west-striking left-lateral faults in the Mina Deflection and the northwest-striking right-lateral faults in the Eastern Domain (Pierce, 2019) of the Central Walker Lane (Figure 1).

In addition to broad structural complexity, the Central Walker Lane has a long history of prolific, moderate-magnitude earthquake swarms and mainshock-aftershock sequences, many of which rupture complex fault geometries. We discuss several of these notable sequences that have occurred near the Monte Cristo Range sequence in the last 40 years (labeled in Figure 1). In May 1980, seismic unrest near the Long Valley Caldera peaked when four $M_w$ 6+ earthquakes occurred on orthogonal and *en echelon* strike-slip faults (Hill, 2006). In 1984, the $M_w$ 5.8 Round Valley sequence occurred southeast of Long Valley along two orthogonal strike-slip faults defined by aftershocks within 30 hours of the mainshock (Priestley et al., 1988; Smith and Priestley, 1993). East of Round Valley, an $M_w$ 5.9 left-lateral earthquake was followed 24-hours later by an $M_w$ 6.3 right-lateral earthquake in the 1986 Chalfant, CA sequence (Smith and Priestly, 1988; Smith and Priestley, 2000). On the eastern edge of the Mina Deflection, the 2004 Adobe Hills swarm (Baca *et al.*, 2014); the long-duration 2011 $M_w$ 4.6+ Hawthorne swarm (Smith *et al.*, 2011); the 2016 $M_w$ 5.4+ Nine-Mile Ranch orthogonal multiplet sequence (Hatch *et al.*, 2019); and the 2020 Mono Lake sequence have also occurred. The left-lateral 1997 Fish Lake Valley (Ichinose et al., 2003) and the left-lateral 2013 $M_L$ 5.1 Columbus Salt Marsh mainshock-aftershock sequence occurred south of the Monte Cristo Range sequence (Figure 1). Finally, the 1994 $M_w$ 5.8 Double Springs Flat earthquake near Lake Tahoe (Ichinose et al., 1988) and the orthogonal 2019 $M_w$ 6.4 and $M_w$ 7.1 Ridgecrest earthquakes (Ross et al., 2019; duRoss et al., 2020) are two additional examples of strike-slip conjugate fault systems that produced significant earthquake sequences in the Northern Walker Lane and Southern Walker Lane, respectively. While all are in similar tectonic environments (i.e., accommodating plate boundary shear), the spatiotemporal behavior, fault kinematics, and aftershock productivity of these sequences varies significantly.

The interplay of multiple faults with different senses of slip to accommodate transtension is a common feature of the abundant seismicity of the Walker Lane, and the 2020 Monte Cristo Range sequence is no exception. The sequence produced 336 $M_L ≥ 3$, 42 $M_L ≥ 4$, and 4 $M_L ≥ 5$ aftershocks through 31 Aug. 2020 (Figures 1 - 3). To explore the fault orientations and slip mechanisms involved in the Monte Cristo Range earthquake, we use precise earthquake relocations in combination with moment tensors for 128 of the largest events. Earthquake relocations are necessary to identify fault geometry at depth, which in recent years has been shown to be an important and complex aspect of strike-slip earthquakes in the western US (e.g., Ross et al., 2019). Precise relocations also provide information on seismogenic depth (Ruhl, Seaman, *et al.,* 2016), which - when interpreted as a locking depth - becomes a primary input to geodetic modeling of regional strain (*e.g.,* Bormann *et al.,* 2016).

Here, we present observations from the aftershock sequence that reveal a network of complex fault geometries that likely ruptured during the complex Monte Cristo Range earthquake. We show that aftershocks occurred on numerous distinct fault structures with a combination of right-lateral, left-lateral, and normal fault motion using moment tensors and earthquake relocations (Figure 2). We compare seismicity data to surface rupture and fracture zone locations (Koehler et al., 2021) and a simple geodetic fault model (Hammond et al., 2020) and find good agreement at depth. Together, these three independent datasets shed light on the complexity of this M6.5 Walker Lane earthquake.

**OVERVIEW OF THE M6.5 MONTE CRISTO RANGE EARTHQUAKE SEQUENCE**

The Monte Cristo Range mainshock earthquake has a w-phase moment magnitude ($M_{WW}$) of 6.5 (USGS Event Page, https://earthquake.usgs.gov/earthquakes/eventpage/nn00725272), full-waveform moment magnitude ($M_w$; this study) of 6.4, and local magnitude ($M_L$) of 6.5 (NSL Event Page, http://www.seismo.unr.edu/Events/main.php?evid=725272; see Figure 3a). We determine the best-fit regional, long-period, full-waveform moment tensors for 128 of the largest events ($M_w$ 3.0–6.4) by assessing a set of trial solutions using 1-km depth increments for highest percent double couple and lowest variance reduction between synthetic and filtered waveforms (Ichinose et al., 2003; see Supporting Text for details). We apply a bandpass filter with a lower frequency band of 0.01 to 0.03 Hz and high frequency limit of 0.1 Hz to broadband waveforms to avoid finite faulting effects. We invert for a moment tensor solution for the mainshock with only five well distributed stations (Figure S1) that predicts the observations at 40 other stations within 500 km (Figure S2). We typically use 2 to 10 stations for the aftershock moment tensors depending on magnitude and noise levels (e.g., Figure S3). We were not able to obtain moment tensor solutions from aftershocks within two hours after the mainshock due to interference from the mainshock coda. Resulting deviatoric moment tensors are shown in Figure 1 for the mainshock and for all events in Figures 2, 4, and 5. While the USGS solution for the mainshock is only 68% double-couple, the NSL moment tensor solution and our solution are both above 95% double-couple; this difference could be due to noise in teleseismic waveforms used in w-phase inversion. Based on our regional moment tensor results and the NSL moment tensor results there is no evidence of large non-double couple components for the mainshock; it was likely a strongly tectonic double-couple earthquake.

Moment tensors are largely strike-slip faulting mechanisms (84%) with some normal faulting (16%) concentrated on the east and west ends of the aftershock zone (Figure 2). Developing first-motion focal mechanisms for many smaller magnitude events using the local, temporary seismic stations may affect this distribution and should be done in future work. Nonetheless, the mixture of normal and strike-slip focal mechanisms is common across transtensional environments like the Walker Lane (see, e.g., Ruhl, Seaman, et al., 2016) and in individual Walker Lane seismic sequences (see, e,g., Ruhl, Abercrombie, et al., 2016, Hatch et al., 2020). Stress axes orientations obtained from a stress inversion (MSATSI, Martinez-Garzon et al., 2014; Hardebeck and Michael, 2006; Lund and Townend, 2007) of our moment tensors are consistent with west-northwest extensional Walker Lane seismotectonics (Ichinose et al., 2003; Figure 2 inset).

We compare moment and local magnitude estimates for the largest events in Figure 3a. Overall, we find that local magnitude estimates are systematically higher than moment magnitudes. We estimate the linear relationship between moment and local magnitude as $M_w$ = $0.77M_L + 0.68$ and use it to convert local magnitudes of smaller events to moment magnitude. The cumulative moment magnitude of aftershocks with moment tensor solutions through 31 Aug. 2020 is $M_w$ 5.35, significantly smaller than the moment magnitude of the mainshock (Figure 3b). The sequence is ongoing and several additional $M_w$>5.0 aftershocks have occurred since 31 Aug 2020.

The magnitude of completeness for the NSL earthquake catalog for this sequence is estimated as $M_L$ 2.2 for analyst-reviewed aftershocks and $M_L$ 1.1 for both reviewed and automatically located events (Figures 3b & 3c). Because the automatic local magnitudes use automatically picked P-wave amplitudes, it is possible that the automatic magnitudes are inflated slightly and that the completeness value is actually lower due to the inclusion of some mis-picked S-waves. The maximum likelihood *b*-value for the sequence to date is 0.87 ± 0.01 for automatic and analyst reviewed earthquakes above the magnitude of completeness and below $M_L$ 5.0. The b-value is 0.83 ± 0.03 for analyst-reviewed events only (Figure 3c). While typical tectonic seismicity has b-values close to 1 (e.g., King, 1983), these values are similar to what is found in other sequences occurring in the Walker Lane. For example, the 2008 $M_L$ 5.1 Mogul, NV earthquake sequence in the Northern Walker Lane had a b-value of 0.94 (Ruhl, Abercrombie, et al., 2016) and estimates for the b-value of the 2019 Ridgecrest sequence has been reported as low as 0.74 (Hauksson et al., 2020). Estimates lower than 1 suggest either underproduction or underdetection of small magnitude earthquakes or higher-than-average production of relatively larger magnitude aftershocks (e.g., Hauksson et al., 2020).

While no closely located foreshocks were observed (Figure S4), the nearby Apr. 2020 $M_w$ 5.2 Mono Lake sequence preceded the Monte Cristo Range earthquake by less than one month (Figure 1). The $M_{ww}$ 6.5 event appears to be the mainshock of a prolific, ongoing aftershock sequence (Figure 3b). The aftershock sequence obeys Båth's law (Mdiff = 1.4 units; Båth, 1965; Helmstetter & Sornette, 2003) and has high aftershock productivity, decaying over time following the modified Omori's law (Utsu, 1961) with p = 0.8 (Figure 3d). Because the real-time telemetry of some of the temporary stations was delayed, there is an artificial increase in daily number of earthquakes at ~10 days and we therefore excluded data in the first 10 days from the Omori's law fit (Figure 3d; Bormann et al., 2021).

The mainshock hypocenter depth is poorly resolved, but likely initiated at a relatively shallow depth near the intersection between the Eastern Columbus Salt Marsh fault zone and the eastern extent of the Candelaria fault (Figure 2). The NSL reviewed catalog depth at the time of this study was 4.8 ± 1.5 km relative to sea level, while our relocated depth estimate is 3.74 km relative to mean regional station elevation of 1.583 km (see next section for relocation details). The moment tensor solution suggests a slightly deeper centroid depth of 8.0 km below sea level (Figure S1). The 073° left-lateral nodal plane of the moment tensor solution and the general east-northeast trend of the aftershocks indicates left-lateral strike-slip motion. The 174° auxiliary nodal plane is similar to the northwest-southeast oriented Eastern Columbus Salt Marsh fault zone, which intersects and offsets the aftershock zone near the mainshock epicenter. Numerous surface ruptures and fracture zones span the seismicity cloud in a complex pattern, suggesting that the mainshock did not occur on a simple planar left-lateral fault but rather on a complex network of faults (Koehler et al., 2021). In the rest of this paper, we will use earthquake relocations and moment tensors to discuss the complex fault structures observed at depth before returning to the complexity of the mainshock rupture.

**ABSOLUTE AND RELATIVE RELOCATION OF THE AFTERSHOCK SEQUENCE**

We select 18,256 events located by NSL from 1 Jan. 2020 to 31 Aug. 2020 and access P- and S-phase arrival data directly from the NSL database (see Data and Resources). Using USGS program HYPOINVERSE (Klein, 1978), we calculate precise absolute relocations (Figures S5 and S6) by applying datum and station corrections (Figure S7) to the entire sequence following Ruhl, Seaman, et al. (2016) and Ruhl, Abercrombie, et al. (2016); please see Supporting Text for more details on the absolute relocation process. This method decreases the median absolute horizontal and vertical location uncertainties from 1.15 and 1.23 km in the NSL catalog to 0.73 and 0.89 km, respectively (Figure S8). We modified the four shallowest layers of an NSL preferred velocity model developed for the Northern Walker Lane (Ruhl, Seaman, et al., 2016) to include slower velocities, which reduced the number of events locating above the surface (Table S1).

To sharpen features seen in the absolute locations, we calculate waveform-based double-difference relocations. We download three-component waveform data from the IRIS Data Management Center (see Data and Resources) for all seismic stations within 150 km of each event, including waveforms for eight temporary stations rapidly deployed by NSL between 16 - 18 May 2020 (see Bormann et al., 2021). We use the double-difference relocation algorithm HypoDD (Waldhauser and Ellsworth, 2000) to pair earthquakes with 5 or more phases with up to 30 neighbors, thus forming 342,287 event pairs from 18,057 events with a total of 220,874 individual phase picks. For each phase pair, we calculate catalog differential times (2.48 million

P-waves and ~54,000 S-waves) and perform sub-sampled waveform cross-correlations using a magnitude-based bandpass filter on windows of 0.5 s and 1.0 s centered on the arrival for P- and S-waves, respectively. Please see Supporting Text for more information on location uncertainties (Figure S8), filtering, cross-correlation (Figures S9 and S10), and relocation tests (Figure S11). This approach results in ~132,000 and ~22,000 P- and S-phase cross-correlations, respectively, with cross-correlation coefficients greater than or equal to 0.6. Using a total of 2.69 million catalog and cross-correlated differential times recorded on 29 near-source and regional stations, we resolve relative relocations for 16,714 events (95%). Events are discarded if they locate above the surface (airquakes) or lose connection to other events. The relocated catalog (Dataset S1) is shown in Figure 4 and subsequent figures as well as electronic supplement Movie S1.

**COMPLEX FAULT NETWORK: AFTERSHOCKS THROUGH 31 AUG. 2020**

As shown by the catalog locations in Figure 2, the aftershocks extend approximately 35 km along an east-northeast-trending zone that extends eastward from the mapped surface trace of the left-lateral Candelaria fault zone (CFZ) to the southern extension of the right-lateral Petrified Springs fault system (PSFS). If the length of the aftershock zone represents the mainshock fault length, it is 1.5 to 2 times longer than expected for an Mw6.5 earthquake rupture (Wells and Coppersmith, 1994). Our relocations (Figures 4 and 5) show that the aftershock zone is composed of many distinct fault structures with various orientations. The northeast-southwest trending zone is offset near the mainshock hypocenter (Figure 4a) by a dense north-striking zone of seismicity extending from the northern terminus of the extensional Eastern Columbus Salt Marsh fault zone (ECSMF). The relocated mainshock is slightly west of the NSL catalog location, but still occurs near the intersection of the ECSMF and the east-northeast-trending aftershock zone. Earthquakes commonly nucleate and terminate at fault intersections (e.g., King, 1983), and the mainshock occurs between two distinct fault zones. The structures defined by our relocated seismicity show distinctly different patterns to the west and east of the mainshock hypocenter. Accordingly, we describe structures in the western and eastern sections separately below.

**Seismicity Patterns in the Western Section**

In the western section (i.e., west of the ECSMF and mainshock hypocenter), seismicity defines a near-vertical fault at depth which broadens towards the surface in a negative-flower-structure-like network of *en echelon* dipping normal and obliquely-slipping faults (Figure 4c). This interpretation is supported by moment tensors, which are shown as lower-hemisphere in mapview figures and back-hemisphere projections in the cross-sections in Figures 4 and 5. In the fault-parallel cross-section A-A', there is shallow zone with relatively fewer aftershocks (between 10 to 25 km distance and above ~3 km depth in Figure 4b) that may represent an area of high slip in the mainshock, although it is not correlated with left-lateral east-northeast-striking surface ruptures. Dense seismicity and pure left-lateral moment tensors cluster deeper than and to the southwest of the mainshock hypocenter (Figure 4b). We infer that this densely-populated vertical fault section is the primary left-lateral fault plane that slipped during the mainshock (Figures 4a-c, 5d-e). The west-dipping ECSMF structure is below the mainshock in oblique cross-sections in Figures 4b and 4d. Only one moderate magnitude aftershock occurred on this structure, but it is well defined by smaller magnitude seismicity. Inclusion of first-motion focal mechanisms for smaller events will improve the kinematic interpretation of fault structures identified via seismicity lineaments.

The narrow near-vertical, left-lateral fault zone extends from a approximately 12 km to 6 km depth, and strike-slip moment tensors align with the geodetically-derived fault plane at these depths (Figures 4c, 5d-e). Above 6 km depth, seismicity bifurcates into two seismicity zones highlighted by ellipses oriented at 60°SE and 70°NW in Figures 4c and 5c-e. Surface rupture and fracture zone locations are indicated by the blue lines on the surface of cross-sections in Figures 4 and 5. Seismicity zones that shallow towards the northwest consistently project towards surface ruptures identified by Koehler et al. (2021) and Dee et al. (2020); see Figures 4c and 5c-e. The shallowest seismicity appears to abut the surface ruptures, and the strike of the ruptures matches closely the northeast-strike of the nodal planes in the normal and oblique moment tensors (Figure 4). We interpret the seismicity as a broad fault-fracture mesh which extends towards the surface but is not necessarily directly connected to the mapped surface ruptures. Several shallow, northeast-striking planar seismicity clusters in the western section show distinct separation in a right-stepping *en echelon* pattern. This is particularly apparent when seismicity is viewed in three dimensions, and we therefore include a fly-through movie in the electronic supplement (Movie S1). Some of these planar structures appear to be subparallel, but we identify dips towards both the west-northwest and east-southeast (see Figures 4 and 5). Many of the moment tensors associated with these dipping features show both normal and oblique strike-slip motion and we interpret them as a set of discontinuous normal faults that together accommodate left-lateral motion and form the relatively broader east-northeast trending aftershock zone seen in the western section (Figure 4a).

One of the most obvious dipping structures is the nearly north-striking extension of the ECSMF near the mainshock hypocenter. We estimate the dip of this seismicity as approximately 70°W by measuring it on a vertical cross-section perpendicular to the ECSMF strike. All west- and northwest-dipping ellipses in Figures 4 and 5 are oriented at 70°, while the southeast-dipping ellipses are at 60°. Together, the slightly southeast-dipping mainshock fault plane(s) and the west-dipping ECSMF form a southwest-facing wedge that bounds the down-dropped Columbus Salt Marsh tectonic basin (gray dashed line on Figure 3). Shallow seismicity directly above these well-defined structures and within geodetic fault plane boundaries shown in Figure 4b is notably sparse, but shallower events cluster towards the end of the aftershock zone (2-6 km distance in Figure 4b). We observe short S-P times (<0.5 s) for some of these earthquakes on the temporary near-source stations, which suggests that these events may indeed be that shallow.

**Seismicity Patterns in the Eastern Section**

In the eastern section (i.e., east of the mainshock), vertical, moderately dipping, and obliquely-crossing structures are visible in map view and in cross-section (Figure 4). At the eastern edge of the aftershock zone, seismicity trends to the northwest towards the southern extension of the Petrified Springs fault system (Figures 3-5, parallel to cross-section 4e, Figure 4a D-D' ). The northwest-trending zone at the eastern edge appears to be composed of many short, discontinuous faults, suggesting a fault-fracture mesh rather than a well-defined, through-going northwest-trending right-lateral fault (Figure 4e and Figure 5g). Events are concentrated at shallower depths on both the western and eastern edges of the aftershock zone (Figures 4b, 5b, and 5g).

Just east of the mainshock hypocenter, seismicity broadens spatially (Figure 4a) and concentrates on multiple steeply dipping oblique structures at depths between 3 and 10 km (Figure 4a, 4b, and 4d). Between the C-C' and D-D' profile lines (Figure 4a), the seismicity cloud collapses into a narrow (<1 km) near-vertical fault strand with small orthogonal cross-faults (Figure 4a, 4b, 5f). Northwest-striking seismicity lineaments align with right-lateral moment tensor nodal planes and show similar length and orientation to mapped surface rupture zones with right-lateral strike-slip (Figure 4a, vertical ellipse in Figure 4f; Koehler et al., 2021; Dee et al., 2020). The remarkable agreement, and yet kinematic inconsistency, between mapped surface ruptures, seismicity trends, and moment tensor lineaments suggests that the Monte Cristo Range mainshock ruptured multiple faults with different senses of slip to accommodate overall left-lateral shear.

**MAINSHOCK RUPTURE: COMPLEX FAULT(S) WITH POSSIBLE SUBEVENTS**

Waveforms of the mainshock are complex, with phases indicating two overlapping earthquakes or sub-ruptures (Figure 6). In the raw waveforms, there appears to be a smaller initial earthquake followed by a larger earthquake 2 - 3 s later. This observation is supported by the automatic source time function generated by the Institut de Physique du Globe de Paris (IPGP; Figure 6 inset; Vallée et al., 2011; Vallee et al., 2013). The IPGP teleseismic body-wave moment tensor has a half-duration of 4.2 s, typical of an M6.5 earthquake, but for the Monte Cristo Range earthquake source time function, the centroid minus hypocenter time is 10.0 s - more than twice the half-duration. The IPGP source time function shows a smaller and shorter initial pulse followed by a larger and longer pulse approximately 3 s after initiation (Figure 6 inset). This further supports that the Monte Cristo Range earthquake was a complex, multiple subevent rupture.

To further explore the hypothesis of complex multi-fault rupture during the mainshock, we discuss the spatiotemporal distribution of aftershocks in relation to the mainshock. The aftershock distribution extends bilaterally from the mainshock hypocenter extending to the full 35-km length of the rupture zone within one day of the mainshock (Figure 5). In cross-section A-A' of Figure 4 and discussed in the previous sections, seismicity concentrates on numerous faults and leaves large seismicity voids. For example, there is a decrease in earthquake density where the aftershock zone crosses the region between the southern Benton Springs fault and the northern Eastern Columbus Salt Marsh fault (approximately 18 km distance in Figure 4b).

The majority of large aftershocks, both east and west of the mainshock, occur within the cross-sectional area of the simple fault model independently developed by Hammond et al. (2020) using geodetic surface displacements (green rectangle in Figure 4b). Large magnitude aftershocks (those with moment tensors) cluster on the deeper vertical fault in the western section and at similar shallow depths as the mainshock in the eastern section (Figure 4b). To show this, we calculate cumulative number and cumulative moment release density plots of aftershocks along the entire fault zone (Figure 7). We bin the relocated hypocenters spatially and in depth using nonoverlapping 2.5 km-by-2.5 km bins which is roughly the size of an average $M_w5$ earthquake (i.e., the approximate magnitude of the largest aftershocks). We use moment magnitudes from the moment tensor inversion when available and convert local magnitudes of smaller earthquakes to moment using the empirical equation derived from earthquakes with moment tensors (see Figure 3). We observe an anticorrelation between the number of aftershocks and the cumulative moment of aftershocks spatially across this zone. Dashed boxes in Figure 7d, marking the regions with the highest number of events from Figure 7c, show that the greatest moment release is not coincident with the greatest number of events. Significantly more earthquakes occurred in the western section, but the cumulative moment release of all $M_w$ ≥ 3.0 earthquakes is concentrated in the eastern section (Figures 4 and 7). This relationship may be related to heterogeneous mainshock slip across the fault zone, preexisting fault heterogeneity, and/or secondary postseismic processes like fluid migration and aseismic slip.

Finally, we discuss the depth distribution of seismicity as it relates to the mainshock rupture. There is weak evidence for downward migration of seismicity through time (Figure 5), but hypocentral depth uncertainties are higher for aftershocks in the first few days of the sequence before the deployment of near-source temporary stations (Bormann et al., 2021). We split the data into half-week (3.5-day) bins and plot the distribution of events in 1-km depth bins through time (Figure 8). Shallow events persist, even after the installation of temporary stations (Figure 8a). The majority of events occur between 3 and 8 km depth (Figure 8c), and the largest magnitude events occur primarily above 8 km depth (Figure 8b).

**DISCUSSION ON COMPLEX EARTHQUAKES**

Ruptures spanning multiple faults have become an increasingly common observation of moderate magnitude continental earthquakes. In the western US, complex ruptures are dominated by M5-7 strike-slip earthquakes rupturing immature and often discontinuous fault zones in the diffuse parts of the plate boundary system (i.e, the Eastern California Shear Zone and the Walker Lane). We introduced many of these examples which occurred near the Monte

Cristo Range earthquake at the start of this paper, but larger magnitude examples also exist. One of the classic examples is the 1992 $M_w$ 7.3 Landers earthquake rupturing four *en echelon,* right-stepping, right-lateral faults in the Eastern California Shear Zone (Haukkson et al., 1993). More recently, the 2019 $M_w$ 6.4 and $M_w$ 7.1 Ridgecrest earthquakes ruptured an unmapped network of orthogonal strike-slip faults in the Southern Walker Lane (DuRoss et al., 2020). Numerous orthogonal fault structures were shown to persist through the entire seismogenic depth range (Ross et al., 2019). Likewise, some of the orthogonal structures we observe in the Monte Cristo Range aftershocks also seem to persist through the seismogenic zone (e.g., Figure 5f). Most of the surface ruptures and seismicity lineaments occur beyond the mapped surface faults included in the USGS Quaternary Faults and Folds database. Earthquakes like these emphasize the need for including smaller faults in seismic hazard analyses and for improving our understanding of how small faults with different slip orientations can cooperate during large, surface-rupturing earthquakes.

Many international examples of complex ruptures also exist. In southwestern China, several shallow moderate magnitude earthquakes have been shown to rupture complex, unmapped strike-slip faults (e.g., 2014 M6.6 Jinggui Earthquake; Wang et al., 2018). In central Italy, seismicity and rupture modeling support rupture of complex multiple-fault geometries (e.g., 30 Oct. 2016 M6.5 Central Italy Earthquake; Scognamiglio et al., 2018). Another interesting, and larger magnitude, example is the surface-rupturing 2016 Mw7.8 Kaikoura, New Zealand earthquake that activated more than 21 faults in a complex transpressional tectonic transition zone between the Hikurangi subduction zone and the right-lateral Alpine fault (Xu et al., 2018). Many of these ruptures showed significant strike-slip and it was not initially clear if the underlying subduction plate boundary had ruptured; inconsistencies existed between geodetic deformation, surface fault offsets, seismological observations of the mainshock, and subsurface seismicity locations including the mainshock hypocenter being significantly offset from the moment centroid (Furlong and Herman, 2017). By incorporating these data sets along with tsunami propagation modeling, it was shown that the mainshock involved synchronous slip on multiple upper crustal strike-slip faults as well as slip on the plate interface at depth (Bai et al., 2017; Furlong and Herman, 2017). Without additional constraints on fault slip at depth, it was difficult to reconcile seismogenic observations with the distributed deformation observed at the surface. These examples highlight the need for using multiple surface and subsurface observations to model complex earthquakes and to assess seismic hazard.

By considering multiple observations of the Monte Cristo Range earthquake and its aftershocks, a model for the complex kinematic relationship between surface ruptures and left-lateral slip at depth can be developed. The left-lateral strike-slip moment tensor of the $M_{ww}$ 6.5 mainshock is well-supported by the geodetic fault model by Hammond et al. (2020) as well as our earthquake relocations and moment tensors below approximately 4 km depth. Above this depth, seismicity separates into numerous distinct fault structures at various depths which may help explain the complexity of observed surface ruptures (Koehler et al., 2021; Dee et al., 2020). The Monte Cristo Range aftershock relocations can be improved; for example, by developing and including a more accurate and detailed velocity model rather than our simple 1D model, or by including analyst-reviewed phase picks rather than the mostly automatic picks used here. More detailed spatiotemporal and fault kinematic analysis is needed to truly understand the kinematics and timing of how these sub faults interacted during and after the mainshock rupture.

**SUMMARY**

The $M_{WW}$ 6.5 Monte Cristo Range earthquake was followed by a highly productive and slowly decaying aftershock sequence (p=0.8) which included >18,000 autodetected events in 3.5 months. Here, we developed double-difference, waveform-based relative relocations that revealed a complex network of faults, many of which cross the inferred 35-km long east-northeast-striking, left-lateral mainshock rupture. Seismicity lineaments correspond with left-lateral, right-lateral, and normal mechanism nodal planes for moment tensors of 128 of the largest earthquakes. The mainshock occurred near the middle of the aftershock zone at the intersection of two distinct zones of seismicity. In the western section, numerous subparallel, shallow, north-northeast-striking faults form a broad flower-structure-like fault mesh that coalesces at depth into a near-vertical, left-lateral fault. This near-vertical fault extends the Candelaria fault at depths between 6 and 12 km and is likely a significant region of slip in the mainshock. Near the mainshock hypocenter, seismicity clusters on a northeast-striking, west-dipping structure which extends north from the Eastern Columbus Salt Marsh normal fault. This fault intersects the near-vertical, left-lateral, strike-slip fault and together these structures bound the down-dropped Columbus Salt Marsh tectonic basin. East of the mainshock hypocenter, seismicity occurs in a narrow, near-vertical, east-northeast-striking fault zone through to its eastern terminus. The eastern section hosts significantly fewer aftershocks than the western section but has more moment release.

We infer that shallow aftershocks throughout the system highlight fault-fracture meshes that connect mapped surface fault systems to the left-lateral fault at depth. Although these fault meshes extend towards the fresh surface ruptures, they do not necessarily connect to them.

Instead, our results imply that this immature left-lateral fault system does not extend to the surface and therefore left-lateral slip at depth is distributed complexly over a wide area in the top five kilometers above the fault.

Based on our structural interpretation and considering the complex waveforms and double-pulse source time function of the mainshock, it is likely that the M6.5 Monte Cristo Range earthquake simultaneously ruptured multiple faults at different orientations. By considering multiple observations of the Monte Cristo Range earthquake and its aftershocks, a model for the complex kinematic relationship between surface ruptures and left-lateral slip at depth can be developed. Detailed spatiotemporal, fault kinematic, and dynamic rupture analyses are needed to truly understand how these sub faults interacted during and after the mainshock rupture.

**DATA AND RESOURCES**

We downloaded the regional earthquake catalog in Figure 1 through the ANSS Comprehensive Catalog (USGS EHP, 2017). We also accessed USGS products (e.g., DYFI reports and w-phase moment tensor) from the USGS event page for the mainshock (https://earthquake.usgs.gov/earthquakes/eventpage/nn00725272/). The SCARDEC source time function was accessed from the IPGP event page (http://geoscope.ipgp.fr/index.php/en/catalog/earthquake-description?seis=nn00725272). The Jan. 1 to Aug. 31 2020 local event locations and phase data were accessed directly from the Nevada Seismological Laboratory database, and phase picks for catalog events are also available through the ANSS Comprehensive catalog (https://earthquake.usgs.gov/data/comcat/). All Quaternary faults shown in maps are from the USGS Quaternary Faults and Folds Database (USGS, CGS, and NBMG, 2020). Simplified surface fault ruptures and fracture zones were obtained from Seth Dee based on detailed mapping in Dee et al. (2021) and Koehler et al. (2021). The geodetic fault model is from Hammond et al. (2021). We accessed waveforms through IRIS Data Services, specifically the IRIS Data Management Center (DMC), using ObsPy, a python library for seismological analysis (Krischer et al., 2015). IRIS Data Services are funded through the Seismological Facilities for the Advancement of Geoscience (SAGE) Award of the National Science Foundation under Cooperative Support Agreement EAR-1851048. We make our moment tensor and relocated earthquake catalogs available in the electronic supplement (Datasets S1 and S2).

**ACKNOWLEDGEMENTS**

We thank the Nevada Seismological Laboratory's analysts and station deployment team for making this study possible. We thank Seth Dee, Rich Koehler, Bill Hammond, Austin Elliot, Gordon Seitz, and Rachel Abercrombie for helpful discussion of the sequence. Map figures were made using the Generic Mapping Tools v. 4.5.14 (www.soest.hawaii.edu/gmt; last accessed November 2015; Wessel & Smith, 1991; Wessel et al., 2013). Movie S1 was made with Fledermaus interactive 3-D geospatial processing and analysis tool (Quality Positioning Services). NSL regional seismic monitoring and the Monte Cristo Range sequence temporary aftershock monitoring deployment is funded by USGS Cooperative Award #G20AC00038 and a rapid earthquake response data collection supplement. Lawrence Livermore National Laboratory is operated by Lawrence Livermore National Security, LLC, for the U.S. Department of Energy, National Nuclear Security Administration under Contract DE-AC52-07NA27344. Finally, we thank two anonymous reviewers for their helpful comments which improved the manuscript.

*Christine J. Ruhl*

*Department of Geosciences*

*The University of Tulsa*

*Tulsa, Oklahoma, USA*

*cruhl@utulsa.edu*

*Emily A. Morton*

*Jayne M. Bormann*

*Rachel Hatch-Ibarra*

*Kenneth D. Smith*

*Nevada Seismological Laboratory*

*University of Nevada, Reno*

*Reno, Nevada, USA*

*emilymorton@unr.edu*

*jbormann@unr.edu*

*rhatch@unr.edu*

*ken@unr.edu*

*Gene Ichinose*

*Lawrence Livermore National Laboratory*

*Livermore, California, USA*

*ichinose1@llnl.gov*

**LIST OF FIGURES**

**FIGURES**

**Figure 1.** (a) Regional seismotectonic map showing the location of the $M_w$ 6.4 Monte Cristo Range mainshock (yellow star), moment tensor, and aftershock sequence in the context of neighboring faults, recent notable seismic sequences (dashed ellipses with year labeled), and digital seismic stations operating during the Monte Cristo Range sequence (white triangles). The towns of Hawthorne and Tonopah, NV are shown with black diamonds. ANSS Catalog seismicity (M > 3.0) occurring from 1 Jan. 1980 and 31 Aug. 2020 (dots color-coded by time and sized by magnitude) are shown with notable sequences labeled and encircled by dashed black ovals. The 2014 National Seismic Hazard Map fault sources (dark gray lines, Shumway (2019)) are shown. Thick dashed lines demarcate the approximate boundary of the Walker Lane. Dotted black box shows extent of maps in subsequent figures. (b) Inset map shows the study location (blue box) in context of larger Pacific/North American plate boundary in Nevada and California. Abbreviations: SAF - San Andreas fault, SN - Sierra Nevada microplate, WL - Walker Lane, ECZS - Eastern California Shear Zone, and BRP - Basin and Range Province.

**Figure 2.** NSL Catalog location map. Earthquakes are sized and colored by local magnitude. Temporary seismic stations are shown by white triangles with labels. USGS Quaternary Faults and Folds database faults are shown in gray. Relevant fault zones are labeled (CFZ = Candelaria fault zone, ECSMF = Eastern Columbus Salt Marsh fault, PSFS = Petrified Springs fault system, and BSF = Benton Springs Fault). Columbus Salt Marsh is highlighted by white dashed polygon. The moment tensor for the $M_w$ 6.41 mainshock is shown at top left and is overlain by MSATSI stress inversion with bootstraps for 128 $M_w$ 3.0+ moment tensors shown in the map (P-axis is red, T-axis is blue, and null axis is green).

**Figure 3.** Earthquake statistics of the Monte Cristo Range aftershocks. (a) Comparison of NSL local magnitudes ($M_L$) with body-wave-based moment magnitudes ($M_w$) determined from moment tensor solutions. Solid line represents the best-fit relationship between the magnitude types. (b) Earthquakes in the region of the Monte Cristo Range sequence (black circles) since 15 Apr. 2020, 1 month preceding the mainshock, through 31 Aug. 2020, with corresponding local magnitudes. Orange line corresponds to the cumulative aftershock $M_w$ for 128 aftershocks $M_L \geq 3$ with moment tensor solutions, culminating at $M_w$ 5.35. (c) Gutenberg-Richter relation of aftershocks and corresponding b-values for all automatic and analyst-reviewed events, and only analyst-reviewed events. (d) Aftershock productivity and corresponding Omori aftershock decay. The best fitting decay corresponds to a decay parameter p=0.8. Orange line indicates the cumulative number of events through 31 Aug. 2020.

**Figure 4.** High-precision earthquake relocations in map view and cross-section. (a) Map of earthquakes and moment tensors. Events are sized by magnitude and colored by depth. USGS Quaternary faults (gray), generalized surface ruptures and fracture zones (blue), and geodetic fault model (green) are also shown. The mainshock has a bold outline. The geodetic fault model is shown in green and surface ruptures are shown in blue. (b-d) Cross-sections for vertical profile lines shown in (a). Events within 1.0 km and surface ruptures within 2.5 km of profile lines are included in the cross-sections. Fault structures discussed in the text are highlighted by dashed ellipses. Southeast-dipping and northwest-dipping ellipses are oriented at 60° and 70°, respectively.

**Figure 5.** High-precision earthquake relocations in map view and cross-section. (a) Map of earthquakes and moment tensors. Events are sized by magnitude and colored by time in days since the mainshock. USGS Quaternary faults (gray), generalized surface ruptures and fracture zones (blue), and geodetic fault model (green) are also shown. The mainshock has a bold outline. The geodetic fault model is shown in green and surface ruptures are shown in blue. (b-g) Cross-sections for vertical profile lines shown in (a). Events within 1.0 km and surface ruptures within 2.5 km of profile lines are included in the cross-sections. Shallow fault structures discussed in the text are highlighted by dashed ellipses. Southeast-dipping ellipses are oriented at 60°.

**Figure 6. Three component d**isplacement record sections for the mainshock. Filtered waveforms (1 Hz highpass) are colored semi-transparent to show overlapping traces and labeled by corresponding seismic network code and station name. P-wave (blue) and S-wave arrivals (red) are marked on the waveforms. Vertical line at the bottom left of the east-component panel indicates 1 mm of displacement. Corresponding IPGP source time function is inset (top center).

**Figure 7.** Density plots of cumulative number of relocated events (a and b) and cumulative moment release (c and d). Cumulative moment release corresponds to 128 aftershocks $M_L \geq 3$ with moment tensor solutions using their relocated hypocenters. Density maps (a and c) show

USGS Quaternary faults (black lines) and relocated event locations (gray dots). Depth cross-sections (b) and (d) include aftershocks within 10 km of line A-A'. Mainshock hypocenter is indicated by the black star. Dashed lines in cross-section (d) correspond to areas from cross-section (b) that had the highest numbers of aftershocks.

**Figure 8.** Depth histograms of earthquake relocations colored by (a) days since the mainshock in 3.5 day increments (gray lines) with relocation velocity model (Table S1, black line), and by (b,c) local magnitude. (b) Events with magnitudes greater than M3.5 are plotted separately from (c) events with magnitudes less than M3.5 for clarity. Note that the centroid for events larger than M4.5 may be offset from the hypocentral depths shown here.

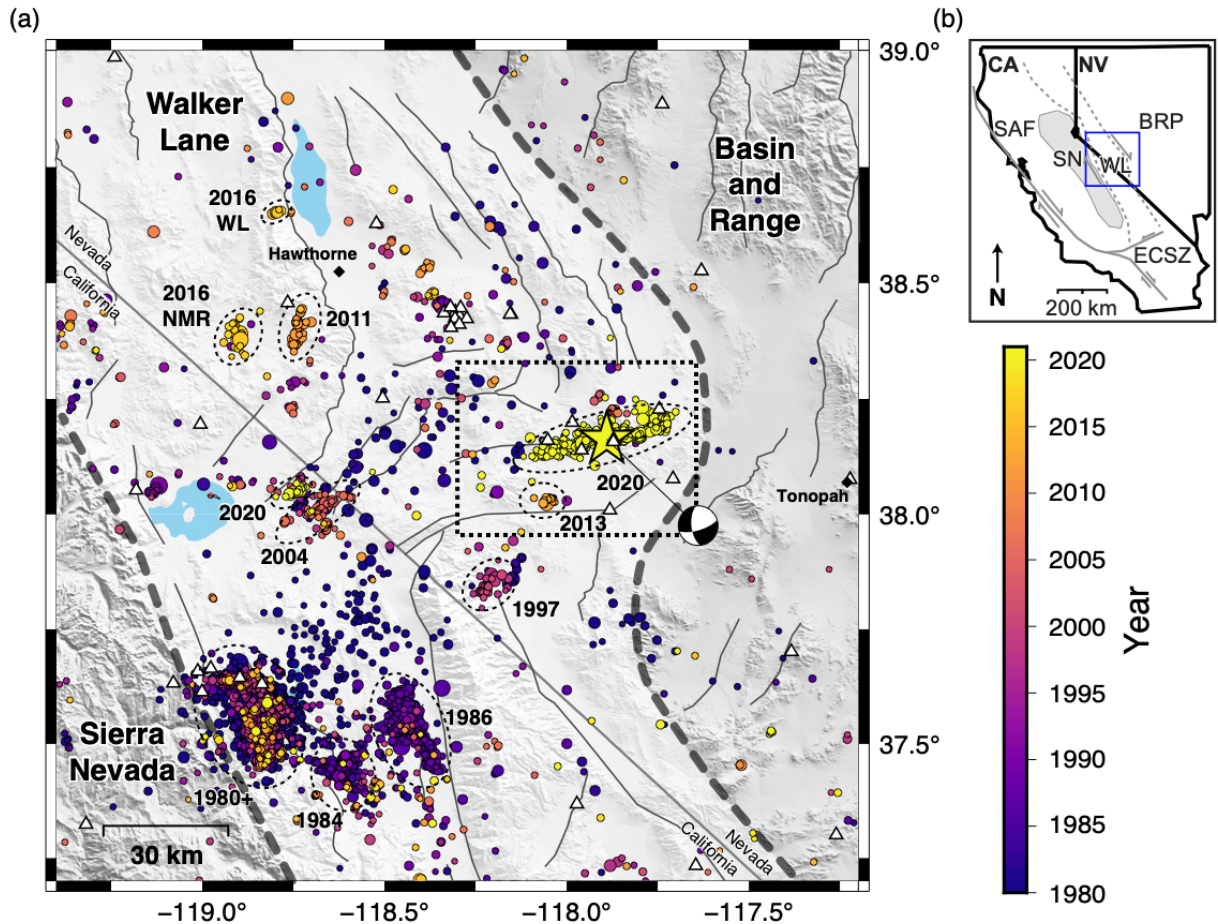

**Figure 1.** (a) Regional seismotectonic map showing the location of the $M_w$ 6.4 Monte Cristo Range mainshock (yellow star), moment tensor, and aftershock sequence in the context of neighboring faults, recent notable seismic sequences (dashed ellipses with year labeled), and digital seismic stations operating during the Monte Cristo Range sequence (white triangles). The towns of Hawthorne and Tonopah, NV are shown with black diamonds. ANSS Catalog seismicity (M > 3.0) occurring from 1 Jan. 1980 and 31 Aug. 2020 (dots color-coded by time and sized by magnitude) are shown with notable sequences labeled and encircled by dashed black ovals. The 2014 National Seismic Hazard Map fault sources (dark gray lines, Shumway (2019)) are shown. Thick dashed lines demarcate the approximate boundary of the Walker Lane. Dotted black box shows

extent of maps in subsequent figures. (b) Inset map shows the study location (blue box) in context of larger Pacific/North American plate boundary in Nevada and California. Abbreviations: SAF - San Andreas fault, SN - Sierra Nevada microplate, WL - Walker Lane, ECZS - Eastern California Shear Zone, and BRP - Basin and Range Province.

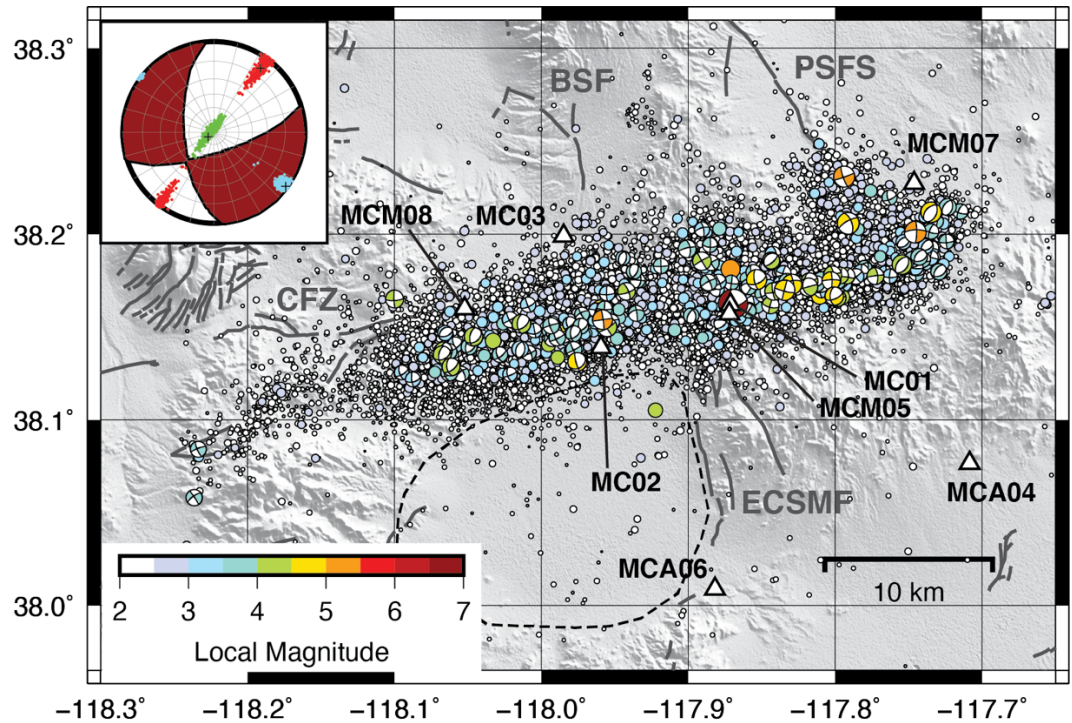

**Figure 2.** NSL Catalog location map. Earthquakes are sized and colored by local magnitude. Temporary seismic stations are shown by white triangles with labels. USGS Quaternary Faults and Folds database faults are shown in gray. Relevant fault zones are labeled (CFZ = Candelaria fault zone, ECSMF = Eastern Columbus Salt Marsh fault, PSFS = Petrified Springs fault system, and BSF = Benton Springs Fault). Columbus Salt Marsh is highlighted by white dashed polygon. The moment tensor for the $M_w$ 6.41 mainshock is shown at top left and is overlain by MSATSI stress inversion with bootstraps for 128 $M_w$ 3.0+ moment tensors shown in the map (P-axis is red, T-axis is blue, and null axis is green).

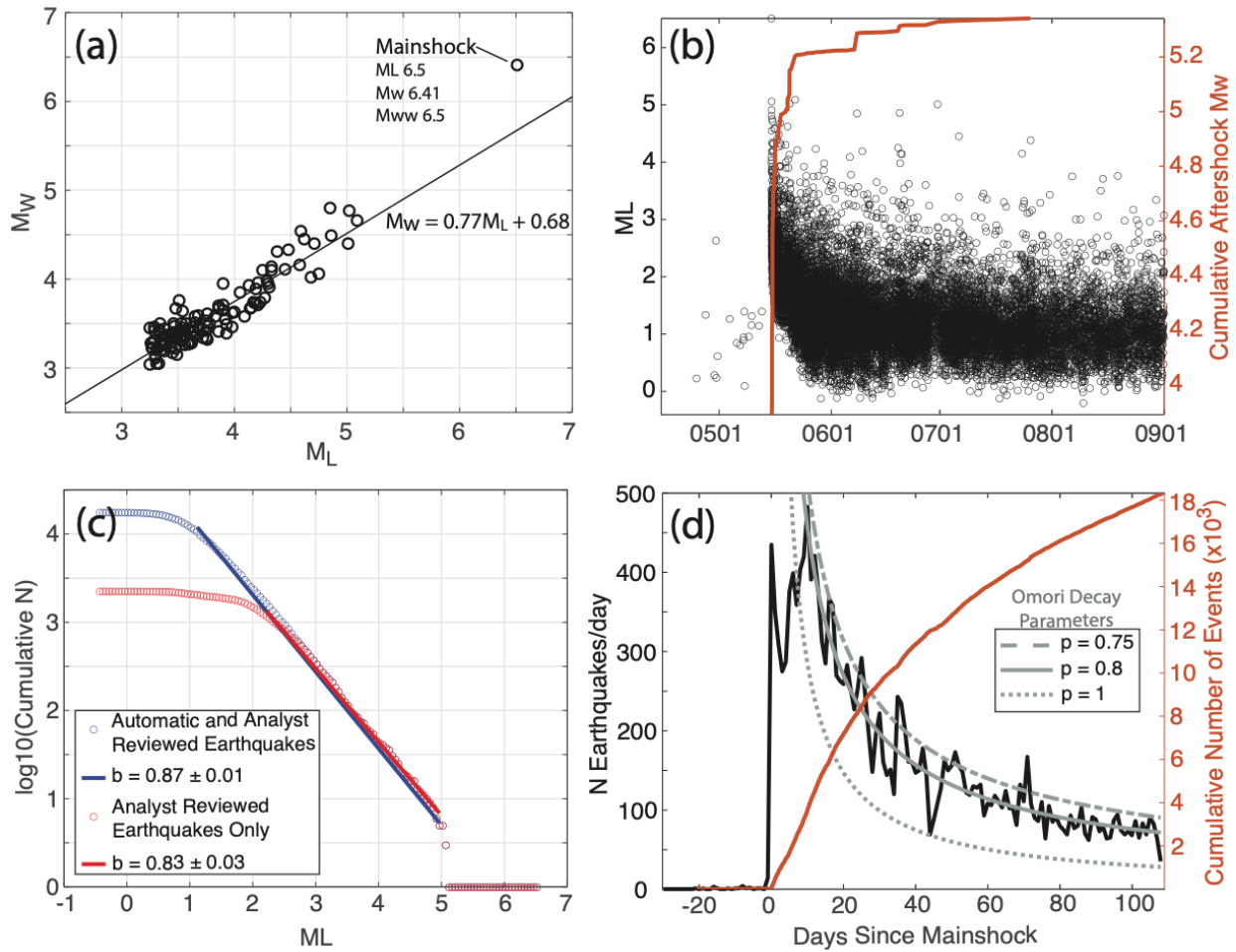

**Figure 3.** Earthquake statistics of the Monte Cristo Range aftershocks. (a) Comparison of NSL local magnitudes ($M_L$) with body-wave-based moment magnitudes ($M_w$) determined from moment tensor solutions. Solid line represents the best-fit relationship between the magnitude types. (b) Earthquakes in the region of the Monte Cristo Range sequence (black circles) since 15 Apr. 2020, 1 month preceding the mainshock, through 31 Aug. 2020, with corresponding local magnitudes. Orange line corresponds to the cumulative aftershock $M_w$ for 128 aftershocks $M_L \geq 3$ with moment tensor solutions, culminating at $M_w$ 5.35. (c) Gutenberg-Richter relation of aftershocks and corresponding b-values for all automatic and analyst-reviewed events, and only analyst-reviewed events. (d) Aftershock productivity and corresponding Omori aftershock decay. The best fitting decay corresponds to a decay parameter p=0.8. Orange line indicates the cumulative

number of events through 31 Aug. 2020.

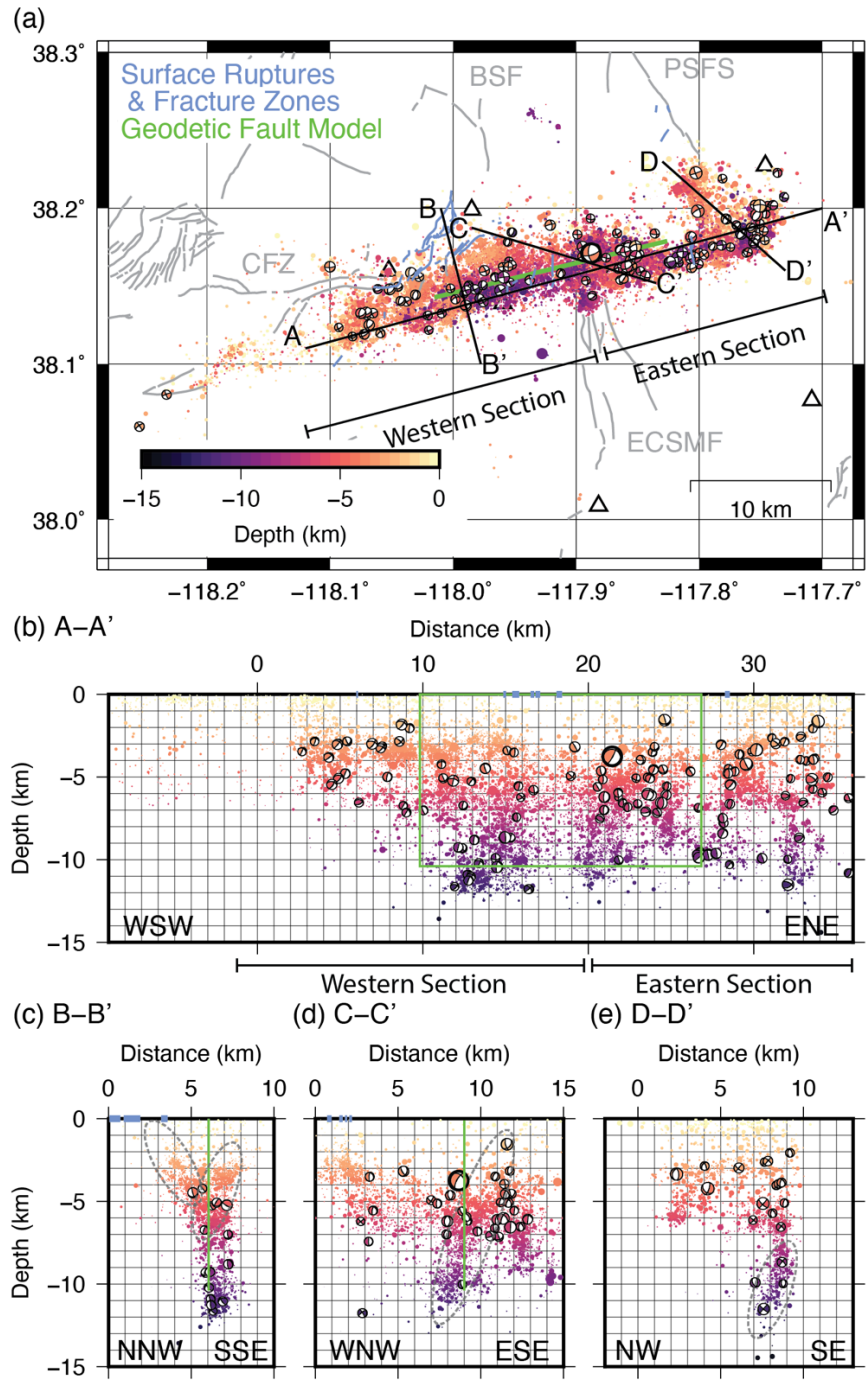

**Figure 4.** High-precision earthquake relocations in map view and cross-section. (a) Map of earthquakes and moment tensors. Events are sized by magnitude and colored by depth. USGS Quaternary faults (gray), generalized surface ruptures and fracture zones (blue), and geodetic fault model (green) are also shown. The mainshock has a bold outline. The geodetic fault model is shown in green and surface ruptures are shown in blue. (b-d) Cross-sections for vertical profile lines shown in (a). Events within 1.0 km and surface ruptures within 2.5 km of profile lines are included in the cross-sections. Fault structures discussed in the text are highlighted by dashed ellipses. Southeast-dipping and northwest-dipping ellipses are oriented at 60° and 70°, respectively.

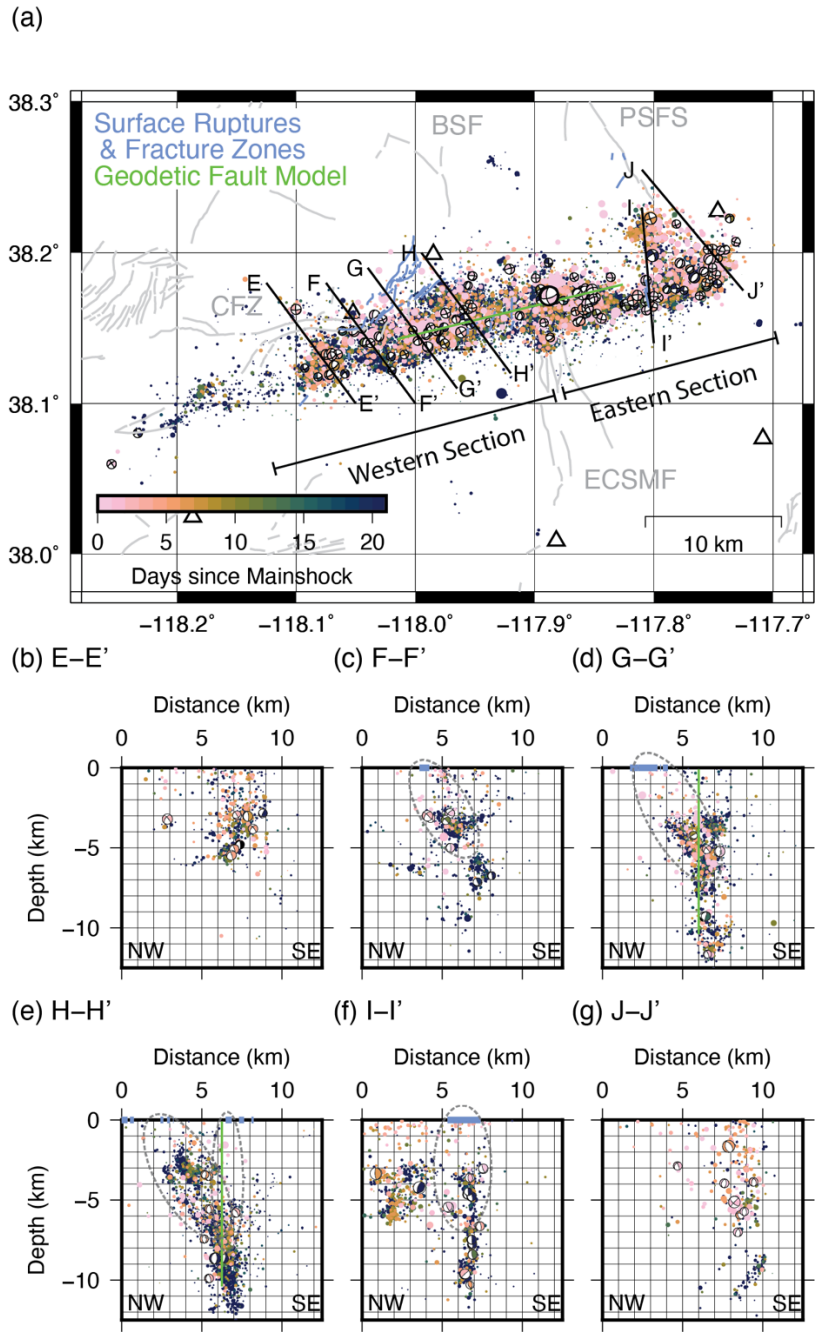

**Figure 5.** High-precision earthquake relocations in map view and cross-section. (a) Map of earthquakes and moment tensors. Events are sized by magnitude and colored by time in days since the mainshock. USGS Quaternary faults (gray), generalized surface ruptures and fracture zones (blue), and geodetic fault model (green) are also shown. The mainshock has a bold outline. The geodetic fault model is shown in green and surface ruptures are shown in blue. (b-g) Cross-

sections for vertical profile lines shown in (a). Events within 1.0 km and surface ruptures within

2.5 km of profile lines are included in the cross-sections. Shallow fault structures discussed in the text are highlighted by dashed ellipses. Southeast-dipping ellipses are oriented at 60°.

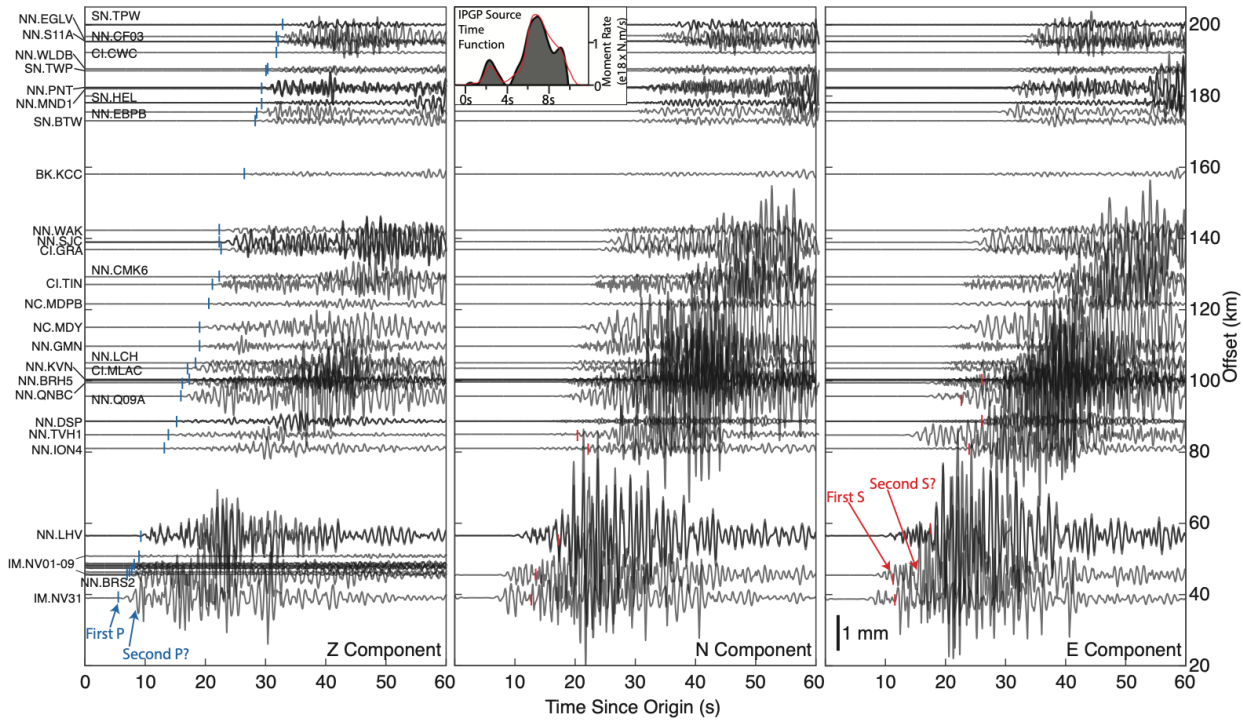

**Figure 6.** Three component displacement record sections for the mainshock. Filtered waveforms

(1 Hz highpass) are colored semi-transparent to show overlapping traces and labeled by

corresponding seismic network code and station name. P-wave (blue) and S-wave arrivals (red)

are marked on the waveforms. Vertical line at the bottom left of the east-component panel indicates 1 mm of displacement. Corresponding IPGP source time function is inset (top center).

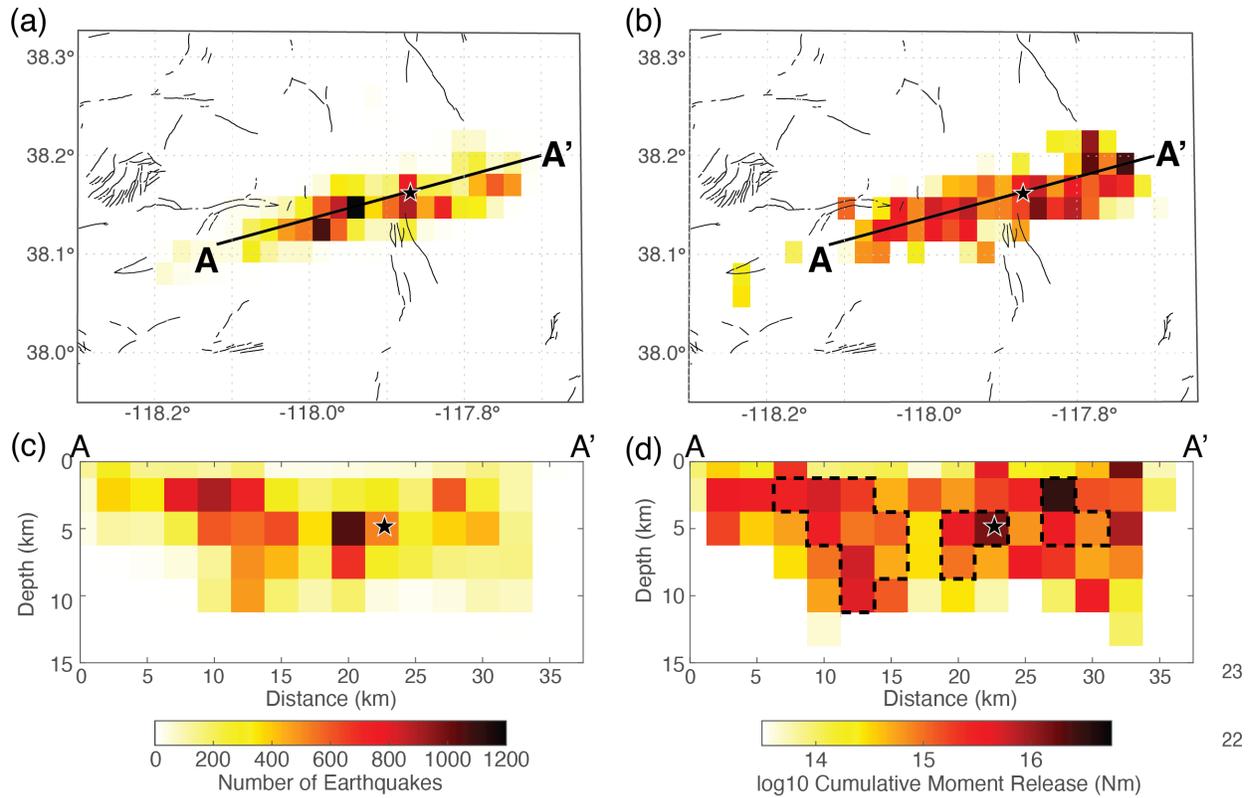

**Figure 7.** Density plots of cumulative number of relocated events (a, c) and cumulative moment release (b, d). Cumulative moment release corresponds to 128 aftershocks $M_L \geq 3$ with moment tensor solutions using their relocated hypocenters. Density maps (a, b) show USGS Quaternary faults (black lines) and relocated event locations (gray dots). Depth cross-sections (c, d) include aftershocks within 10 km of line A-A'. Mainshock hypocenter is indicated by the black star. Dashed lines in cross-section (d) correspond to areas from cross-section (c) that had the highest numbers of aftershocks.

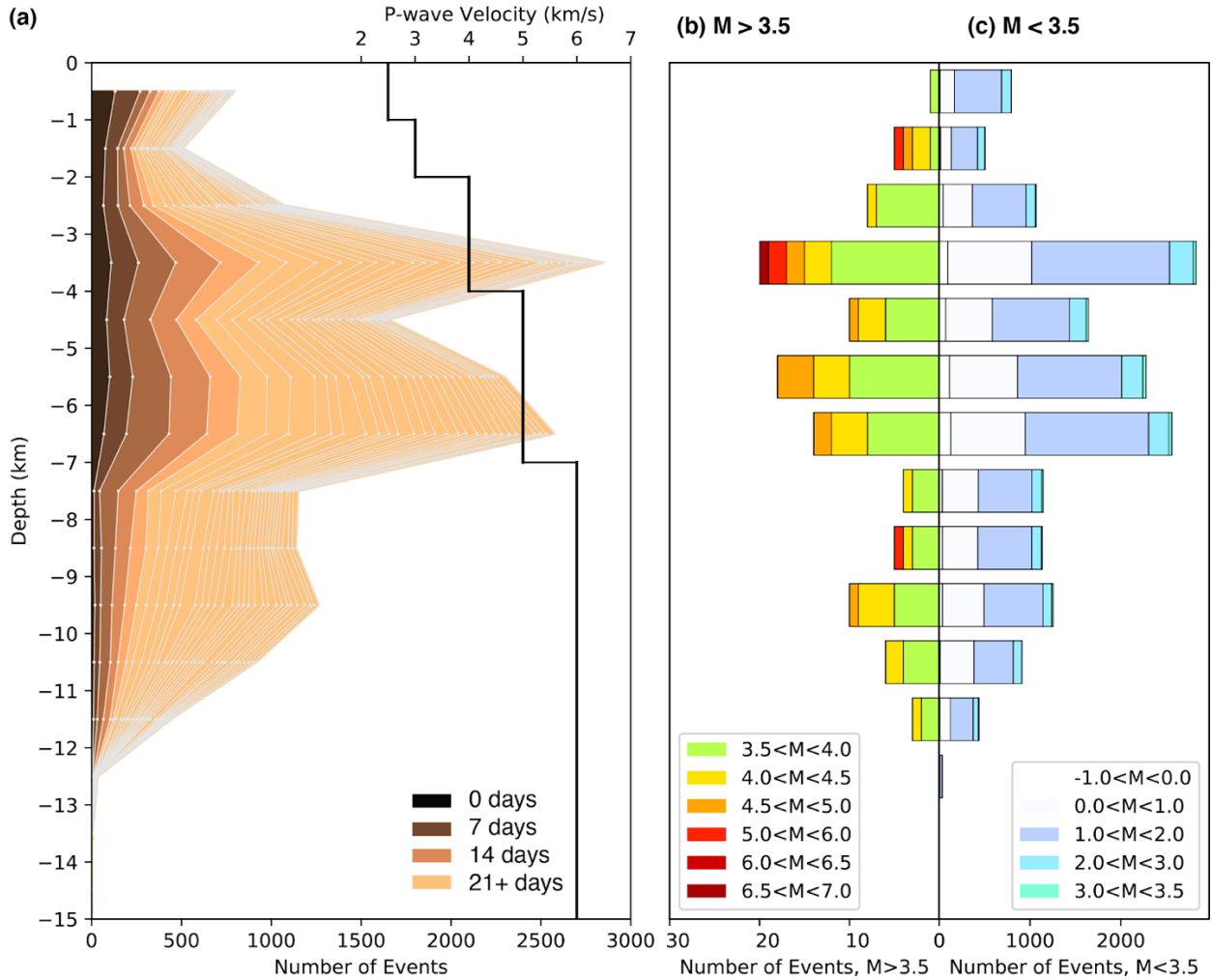

**Figure 8.** Depth histograms of earthquake relocations colored by (a) days since the mainshock in 3.5 day increments (gray lines) with relocation velocity model (Table S1, black line), and by (b,c) local magnitude. (b) Events with magnitudes greater than M3.5 are plotted separately from (c) events with magnitudes less than M3.5 for clarity. Note that the centroid for events larger than M4.5 may be offset from the hypocentral depths shown here.